\documentclass[letterpaper]{article} 
\usepackage{aaai25}  
\usepackage{multirow}
\usepackage{amssymb}
\usepackage{amsmath}
\usepackage{booktabs}
\usepackage{times}  
\usepackage{helvet}  
\usepackage{courier}  
\usepackage[hyphens]{url}  
\usepackage{graphicx} 
\usepackage{natbib}  
\usepackage{caption} 
\usepackage{algorithm}
\usepackage{algorithmic}

\usepackage{newfloat}
\usepackage{listings}
\DeclareCaptionStyle{ruled}{labelfont=normalfont,labelsep=colon,strut=off} 
\floatstyle{ruled}
\newfloat{listing}{tb}{lst}{}
\floatname{listing}{Listing}
\title{Diff-Symbo: Text-Controlled Long-Duration Symbolic Music Generation Using Autoregressive Latent Diffusion Model}
\author{
    Zhiwei Lin\textsuperscript{\rm 1}, Jun Chen\textsuperscript{\rm 1}, Boshi Tang\textsuperscript{\rm 1}, Weihao Wu\textsuperscript{\rm 1}, Jing Yang\textsuperscript{\rm 2}, Yaolong Ju\textsuperscript{\rm 2}, \\
     Fan Fan\textsuperscript{\rm 2}, Zhiyong Wu\textsuperscript{\rm 1 \rm 3} \\
}
\affiliations{
    \textsuperscript{\rm 1}Shenzhen International Graduate School, Tsinghua University, Shenzhen, China\\
    \textsuperscript{\rm 2}Huawei Technologies Co., Ltd., Shenzhen, China\\
    \textsuperscript{\rm 3}The Chinese University of Hong Kong, Hong Kong SAR, China\\

}

\usepackage{bibentry}

\begin{document}

\maketitle

\begin{abstract}
Text-controlled symbolic music generation has recently gained research attention due to its versatile, flexible and straightforward approach to music composition. 
However, previous approaches tend to generate symbolic music with compromising quality, diversity, controllability and limited duration.
In this paper, we present Diff-Symbo, an innovative method that uses latent diffusion model (LDM) to generate high-quality, diverse and long-duration symbolic music.
To address the lack of text-symbolic music dataset, we develop a comprehensive dataset with 19,345 text templates by employing large language model.
Furthermore, we design a music information encoder to reduce the training overhead while extracting more effective control representations.
Given textual descriptions, our proposed method leverages LDM to improve the quality and diversity of music generation. Our method also improves the duration and the compositional consistency of music generation through an autoregressive approach.  
Experimental results show significant improvements of Diff-Symbo in text controllability, duration, and the quality of generated music compared to the baseline models such as GPT-4, MuseCoco and Multitrack Music Transformer (MMT). 
As one of the pioneer models in this field, Diff-Symbo paves the way towards controllable and high-quality symbolic music composition based on LDM, offering valuable contributions to both music amateurs and practitioners.
\end{abstract}

\section{Introduction}

Automatic music generation is an active research field, with a wide range of applications from human-computer interactive composition to video music scoring.
Music can be generated conditioned on modalities such as audio, text, image, and video, among which text-controlled music generation is in particular useful. This is due to the extensive accessibility of texts and its explicit control of music by describing style, genre, and instrumentation specifically, which can be essential for the creativity and personalization of music composition.    
Current research in this field mainly focuses on generating music in the form of audio waveforms, while the generation of \textit{symbolic music} is still in its very early stages.
Compared to audio, music in the symbolic form is represented as a series of notations (e.g., MIDI). This symbolic format allows intricate adjustments on elements such as tempo, pitch, dynamics, and instrumentation with great ease, offering a more versatile and precise approach to music composition. This level of control can be invaluable for composers, musicians, music educators, and those looking to create highly customized, expressive compositions tailored to specific contexts or emotions.

Four factors are critical for text-controlled symbolic music generation: music \textit{quality}, \textit{diversity}, \textit{duration} and text \textit{controllability}. 
Music quality is essential for evaluating music generation models, as it focuses on whether the generated music sounds harmonious and pleasant. 
Diversity refers to the ability of such models to generate music with a wide variety of styles and creative elements.
The duration of music generation directly affects its usability, as overly short pieces may lack coherent music structure and can be difficult to use in practice.
Text controllability means that the generated music conforms to the musical elements given in the textual description, which is crucial for text-controlled symbolic music generation.

For symbolic music generation, current researches primarily rely on Transformer decoder-based architectures (e.g., MuseCoco \cite{lu2023MuseCoco} and MMT \cite{dong2023multitrack}) or diffusion models (e.g., Polyffusion \cite{min2023polyffusion}). Decoder-only models can generate music with longer duration through an autoregressive approach, but they fall short in quality and diversity compared to diffusion models. On the other hand, diffusion models show advantages in music quality and diversity, but they can only generate music with fixed duration, compromising their flexibility and application in various scenarios.

To integrate the advantages of the diffusion model and the autoregressive approach, we introduce Diff-Symbo, a model capable of generating high-quality, diverse, long-duration symbolic music that accurately aligns with text descriptions.
First, we develop a larger and more comprehensive dataset\footnote{https://anonymous.4open.science/r/templates-8DA8/} with 19,345 text templates to overcome the scarcity of datasets.
Then we design a music information encoder (MI Encoder) to reduce the training overhead while extracting more effective control representations.
Moreover, we design a model by utilizing LDM for high-quality and diverse text-controlled symbolic music generation. 
Finally, to enable long-duration and contextually coherent music generation, we introduce musical contextual information to control music generation through an autoregressive approach. 
Subjective and objective experiments\footnote{Demo page: https://apply74.github.io/Diff-symbo/} show that, compared to baselines GPT-4, Musecoco and MMT, Diff-Symbo can generate longer, higher-quality and more diverse symbolic music that better matches textual descriptions. 
With Diff-Symbo, users are able to obtain consistent music up to several minutes from textual descriptions, which greatly improves the efficiency of music creation.

\vspace{-0.1cm}
\section{Relative Work}

\textbf{Text-to-Music Generation} With the recent advancement of Artificial Intelligence Generated Content (AIGC) research, both academia and industry focus on text-to-music generation, where music is represented as audio waveforms. Existing literature in this category often contains three components: 1) A text encoder that maps words into latent space as text embeddings, such as T5 \cite{raffel2020exploring}, CLAP \cite{elizalde2023clap} or MuLan \cite{huang2022mulan}. 2) An autoencoder that condenses waveforms into either discrete acoustic tokens via residual vector quantization \cite{van2017neural, zeghidour2021soundstream, defossez2022high} or continuous acoustic embeddings \cite{schneider2023mo, liu2023audioldm, evans2024fast}. 3) Acoustic modelling that takes in text embeddings and outputs acoustic tokens/embeddings, which are then used by the autoencoder to generate audio waveforms. For example, MusicGen \cite{copet2024simple} and MusicLM \cite{agostinelli2023musiclm} conduct acoustic modelling in an auto-regressive manner, while MAGNET \cite{ziv2024masked} uses masked language modelling, both using the transformer architecture. Alternatively, this process can be achieved with diffusion via the process of noise addition and denoising, such as AudioLDM series \cite{liu2023audioldm, liu2023audioldm2}, Mustango \cite{ melechovsky2023mustango} and Stable Audio series \cite{evans2024fast, evans2024long}. 

However, the music elements in the generated waveforms are implicit and therefore hard to adjust, which limits the practical usability of text-to-waveform generation.
In comparison, symbolic music, represented as a sequence of discrete musical elements, is more amenable to manipulation and modification, which allows adjustments at both low and high levels. 

Research on text-to-symbol music generation is currently in the preliminary exploration stage. 
BUTTER \cite{zhang2020butter}, a music-sentence representation framework, is designed with the disentanglement of music representations based on VAE and cross-modal alignment. It can generate music in ABC notation from texts but such descriptions are constrained to cover three exact keywords.
The advanced language model GPT-4 \cite{achiam2023gpt} can also generate ABC notation music given prompts, but it lacks any significant form of harmony, primarily because it is designed to understand and generate textual data, not specifically music. 
Wu and Sun \cite{wu2022exploring} explored the text-to-music potential of pre-trained language models. Despite fine-tuned on over 200k text-ABC notation music pairs, their model struggles to accurately match the musical attributes in texts and the generated music is limited to solo tracks.
MuseCoco \cite{lu2023MuseCoco} divides text-to-symbolic music generation into a two-stage task. 1) Text-to-attributes understanding: Fine-tuning the pre-trained model BERT \cite{devlin2018BERT} to predict musical attributes in text descriptions. 2) Attributes-to-music generation: Using musical attributes as prefix tokens to autoregressively generate music with Transformer decoder-only architecture. MuseCoco can generate symbolic music that matches text descriptions. However, the length of generated music is limited and the training cost is considerable.

\textbf{Diffusion Model for Music Generation} 
Lately, there have been multiple efforts to incorporate diffusion models \cite{ho2020denoising} into symbolic music generation.
Polyffusion \cite{min2023polyffusion} applies the diffusion model to generate polyphonic music scores by regarding music as image like piano roll representations.
Wang et al. \cite{wang2024whole} utilizes a cascaded diffusion model to generate whole-song symbolic. 
Additionally, the Latent Diffusion Model (LDM) implements diffusion in the latent space produced by modality encoders \cite{kingma2013auto}.
Due to its satisfying generation performance, LDM has been widely apply in various generative domains, including video \cite{blattmann2023stable,ni2023conditional, bar2024lumiere} , image \cite{muller2023multimodal,podell2023sdxl,rombach2022high} and audio \cite{chen2024musicldm,liu2023audioldm,ghosal2023text}.
The concept of LDM was first introduced and applied to the image domain by Robin et al.\cite{rombach2022high}. Later, AudioLDM \cite{liu2023audioldm} extends the LDM to text-to-audio generation, and is able to generate sound effects and music that conform to textual descriptions.
Gautam et al. \cite{mittal2021symbolic} applied the diffusion model to discrete symbolic music for the first time using the 2-bar MusicVAE \cite{roBERTs2018hierarchical}. 

Existing LDM-based work is unable to generate long-duration symbolic music conditioned on textual descriptions. To solve this issue, we propose a contextual learning strategy that generates each new music segment based on preceding segments in an autoregressive manner. This strategy guarantees that long-term music generation can be consistent in the overall composition style, emotion and instrumentation. 
\vspace{-0.1cm}
\section{Method}
\begin{figure*}[htpb]
  \centering
\includegraphics[width=1.0\linewidth]{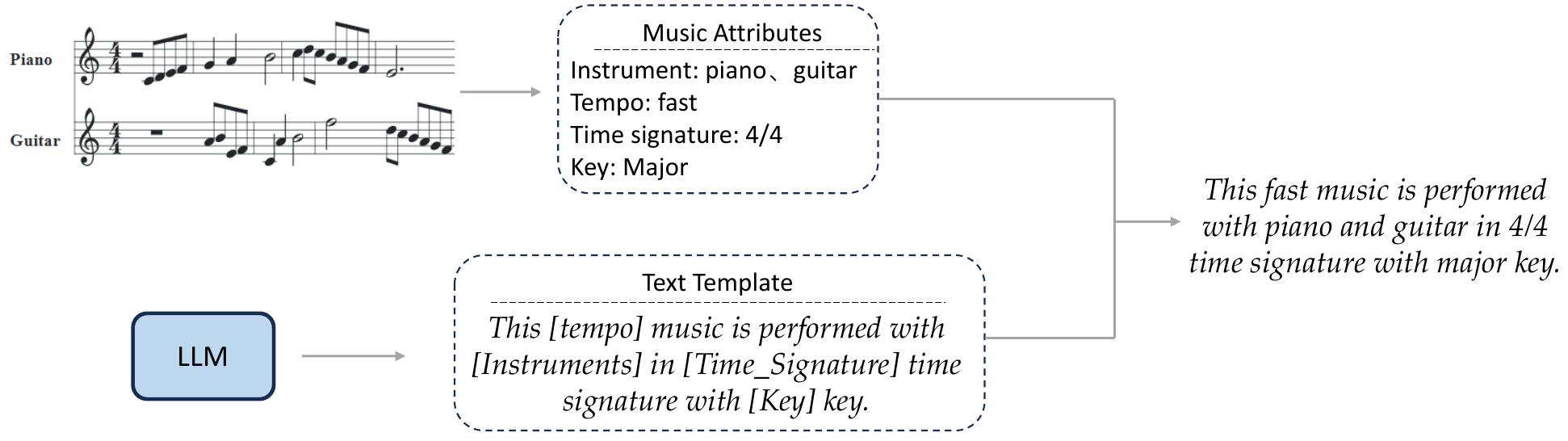}
\vspace{-0.1cm}
  \caption{The pipeline of data construction. We extract the music attributes from MIDI files and use text template given by GPT-4 to generate natural language description.}
\label{fig:data}
\end{figure*}
\begin{figure*}[htpb]
  \centering
\includegraphics[width=0.95\linewidth]{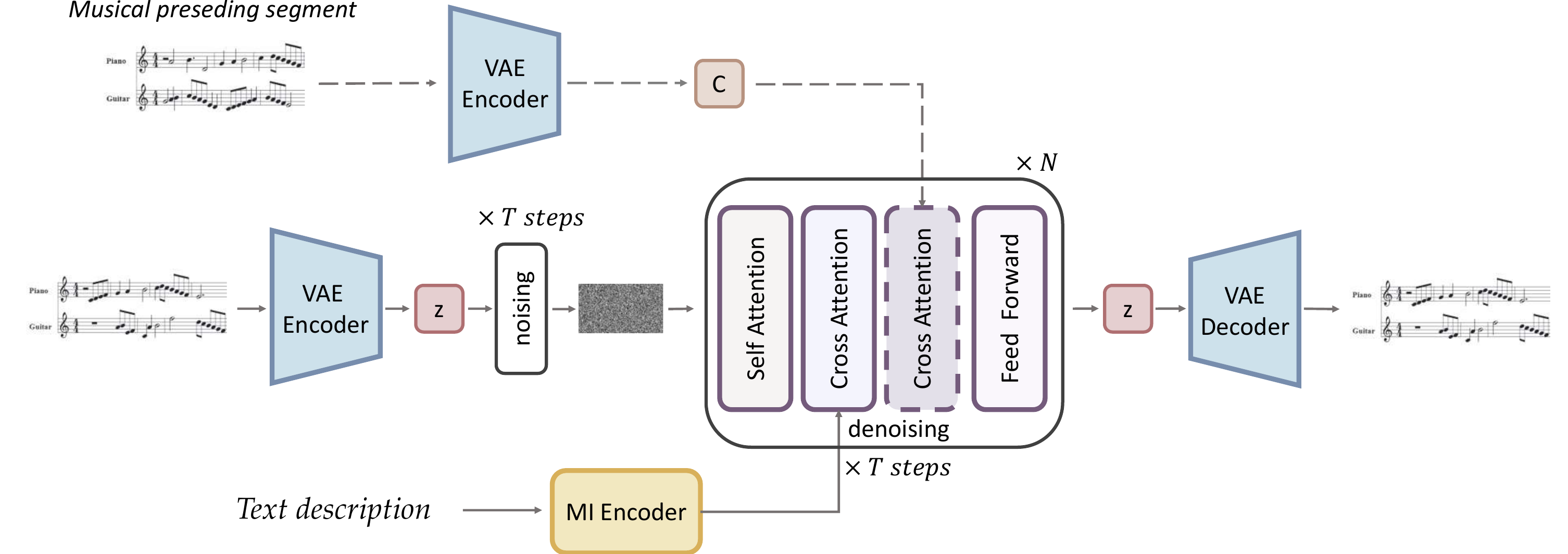}
  \caption{The overall architecture of Diff-Symbo. MI-Encoder represents the music information encoder. The LDM introduces the dummy module for full parameter fine-tuning for long-duration generation.}
\label{fig:overall}
\end{figure*}

Diff-Symbo aims to generate high-quality, diverse, long-duration symbolic music based on natural language text descriptions. The overall architecture of Diff-Symbo, as depicted in Figure \ref{fig:overall}, comprises music information encoder, Multi-view MidiVAE, and Transformer-based diffusion model.
The training process consists of two main stages: LDM training and LDM fine-tuning. 
We first train a LDM to generate symbolic music that matches the text description. 
Subsequently, as indicated by the dashed line in Figure \ref{fig:overall}, we design a musical context module and fine-tune the LDM with all parameters, enabling the model to autoregressively generate long-duration music that is consistent in the overall composition style, emotion and instrumentation. We will provide specific descriptions in the following subsections.

\subsection{Data Construction}
To train Diff-Symbo, a dataset pairing text descriptions with symbolic music is required while existing datasets cannot fulfill our requirement. 
To build such a dataset, we draw inspiration from MuseCoco \cite{lu2023MuseCoco} and create a more extensive and comprehensive text template dataset.
As demonstrated in Table \ref{attribute}, musical attributes can be divided into objective and subjective properties. 
Objective attributes such as instruments and tempo can be directly extracted from symbolic music, while attributes like pitch range require computation through specific rules. 
Subjective attributes must be obtained from pre-existing labels in the dataset.
We design single-sentence templates for each musical attribute. For various combinations of musical attributes, we use GPT-4 \cite{achiam2023gpt} to rewrite these templates to generate 3-6 natural language descriptions with the same meaning.
Considering that arbitrary combinations of musical attributes might occur in practical applications, we apply this process to each combination, ultimately constructing a dataset containing 19,345 text templates that covers all musical attribute combinations. 
As shown in Figure \ref{fig:data}, by extracting musical attributes from symbolic music and using these text description templates, we can convert musical attribute keywords into natural language descriptions, thereby building a dataset that pairs text descriptions with music.

\begin{table*}[htpb]
  \caption{Musical attribute descriptions in Diff-Symbo.}
  \label{attribute}
  \centering
  \begin{tabular}{lllll}
    \toprule
    \cmidrule(r){1-2}
    Type & Attribute & Derivation Method & Description   \\
    \midrule
    \multirow{8}{*}{Objective}  & Instrument & Direct& instruments played in the music   \\
    & Tempo & Direct & the tempo of the music clip \\
    & Time Signature & Direct &  the time signature of the music clip \\
    & Pitch Range & Rule-Based  & the number of octaves   \\
    & Rhythm Danceability & Rule-Based & whether the piece sounds danceable  \\
    & Rhythm Intensity & Rule-Based & the intensity of the rhythm \\ 
    & Key & Rule-Based& the tonality of the music clip \\
    \midrule
    Subjective & Emotion & Label-based & the emotion of the music clip \\
    \bottomrule
  \end{tabular}
\end{table*}

\subsection{Music Information Encoder}
Although pre-trained BERT can extract semantic features from text, the computation cost of subsequent diffusion models increases quadratically with the length of the text description. 
To reduce resource consumption and remove information unrelated to music from semantic representations, we propose the Music Information Encoder, which consists of a frozen
pre-trained BERT, the attention
mechanism and learnable parameters $\{Q_1,Q_2,...,Q_m\}$, where m represents the total number of predefined musical attributes. 
\begin{figure}[htpb]
  \centering
\includegraphics[width=1.0\linewidth]{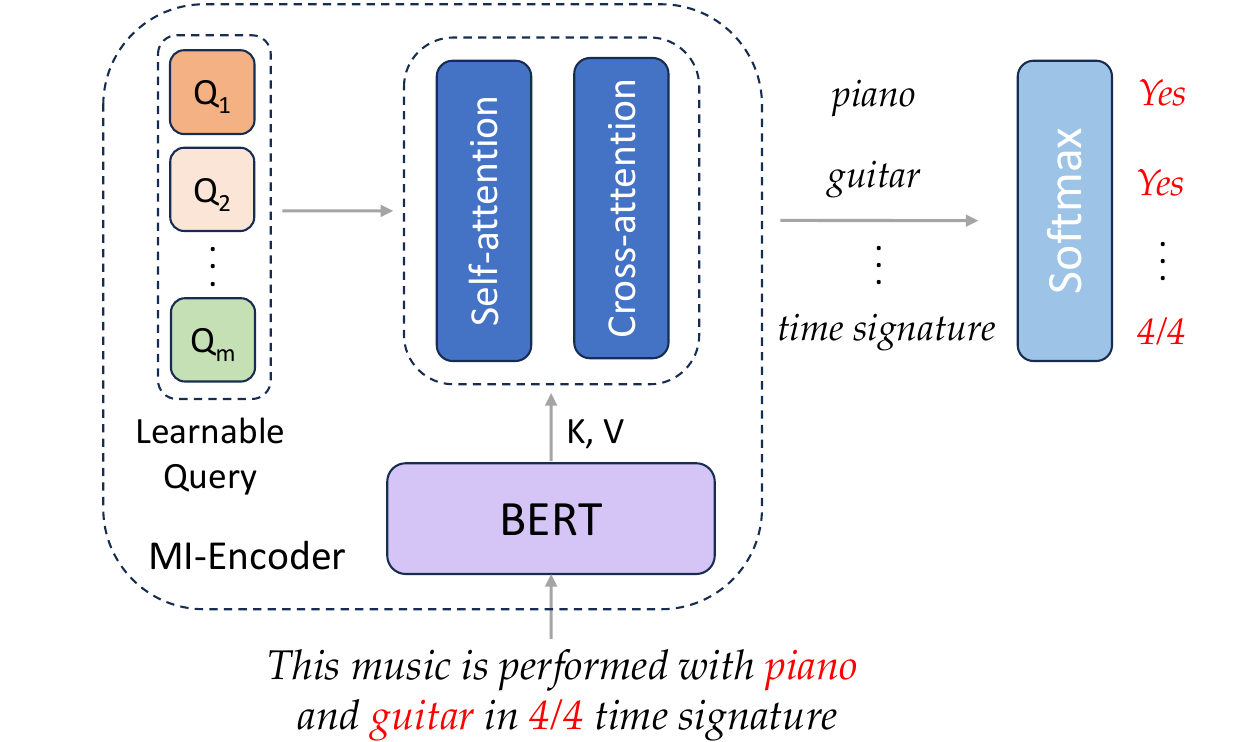}
\vspace{-0.25cm}
  \caption{The architecture of MI-Encoder.}
  \vspace{-0.15cm}
\label{fig:MI-encoder}
\end{figure}
To remove information unrelated to music from semantic, we design a music attributes classification task to train the encoder. 
As shown in Figure \ref{fig:MI-encoder}, these queries are initially processed through the self-attention layer, then used as queries for the cross-attention layer, employing the semantic representations extracted by BERT as keys and values. 
Finally, the output from each query is classified for musical attributes through a softmax classifier, such as the presence of "piano" and "guitar", and the value of time signature.
Notably, during the training of the encoder, we can generate a large number of attribute combinations randomly to create corresponding texts, without the need to specifically extract attributes from MIDI files. 
This approach allows us to construct the training dataset more flexibly and ensures the diversity of data.
After training, we use the output of encoder as the text condition $C_{txt} \in R^{m \times d}$ for latent diffusion model, where d represents the dimension of learnable parameters.

\subsection{LDM for Text-to-Symbolic Music Generation}
LDM is a probabilistic model designed to fit the data distribution $p(x)$ by performing denoising in the latent space. LDM initially encodes the original high-dimensional data $x$ into a lower-dimensional latent representation $z = E(x)$ to enable efficient training.
In text-to-symbolic music generation, our goal is to generate symbolic music $x$ by given text description $y$.

As shown in Figure \ref{fig:overall}, we employ the Multi-view MidiVAE\cite{lin2024multi} to encode symbolic music into the latent variable $z_0 \in R^{L \times C}$, where $L$ represents the number of bars in the music clip and $C$ represents the channel dimension. 
To enhance the quality and diversity of music generation during inference, we train LDM with classifier free guidance (CFG) \cite{ho2022classifier}, for which we arbitrarily discard text conditions $C_{txt}$ with a fixed probability of 20\% to train the conditional LDM $\epsilon_{\theta}(z_t,t,C_{txt})$ and the unconditional LDM $\epsilon_{\theta}(z_t,t)$.

In the noising process, noise is added to the original data distribution with a fixed schedule $\alpha_1, \alpha_2,..., \alpha_T$, where $T$ is the total number of timesteps, and $\overline{\alpha}_t = \prod_{i=1}^t \alpha_i$. 
The process can be described by the following formula.
\begin{align}
    q(z_t \mid z_{t-1}) &= \mathcal{N}\left(z_t; \sqrt{\alpha_t} z_{t-1}, (1 - \alpha_t)I\right) \\
    q(z_t \mid z_0) &= \mathcal{N}\left(z_t; \sqrt{\overline{\alpha}_t} z_0, (1 - \overline{\alpha}_t)I\right)
\end{align}

The goals of both conditional and unconditional LDM are to mirror score matching by optimizing the denoising objective:
\begin{equation}
    \mathcal{L}_{\text{cond}} = \mathbb{E}_{z_0, t, \epsilon} \left\| \epsilon - \epsilon_{\theta}(z_t, t, C_{txt}) \right\|^2
\end{equation}

\begin{equation}
    \mathcal{L}_{\text{uncond}} = \mathbb{E}_{z_0, t, \epsilon} \left\| \epsilon - \epsilon_{\theta}(z_t, t) \right\|^2
\end{equation}
During inference, we employ the denoising process, starting with 
$z_T$ sampled from $\mathcal{N}\left(0,1\right)$ and a guidance scale $\omega$, to generate the latent of symbolic music conditioned on the given text condition $C_{txt}$,
with the following denoising process:
\begin{align}
    p_{\theta}(z_{t-1} \mid z_t) &= \mathcal{N} \left(z_{t-1}; \mu_{\theta}(z_t, t, C_{txt}), \sigma^2_t I \right) \\
    \mu_{\theta}(z_t, t, C_{txt}) &= \frac{1}{\sqrt{\alpha_t}} \left( z_t - \frac{1 - \alpha_t}{\sqrt{1 - \overline{\alpha}_t}} \tilde{\epsilon}_{\theta}(z_t, t, C_{txt}) \right)\\
    \sigma^2_t &= \frac{1 - \overline{\alpha}_{t-1}}{1 - \overline{\alpha}_t} (1 - \alpha_t)
\end{align}
\begin{equation}
    \tilde{\epsilon}_{\theta}(z_t, t, C_{txt}) \leftarrow \omega \epsilon_{\theta}(z_t, t, C_{txt}) + (1 - \omega) \epsilon_{\theta}(z_t, t)
\end{equation}

After we obtain the music latent variable $z_0$, we use the decoder of Multi-view MidiVAE\cite{lin2024multi} to convert it into a symbolic music piece of 8 bars in length.
Given that music and its latent distribution exhibit distinct temporal relationships, our LDM employs the Transformer Encoder architecture which consists of multiple Transformer blocks to capture the temporal dynamics of music.
\subsection{Autoregressive Approach for Long-duration Music Generation}
For LDM, the length of the generated music will be limited by the VAE, e.g., to generate 64-bar music we need to train a VAE that can encode 64-bar music.
However, excessively long music lengths can lead to computation cost increase and poor VAE reconstruction results.
To address this challenge, Gautam et al. \cite{mittal2021symbolic} divided a 64-bar music composition into 32 2-bar segments, each encoded by a 2-bar MusicVAE\cite{roBERTs2018hierarchical}. These segments were then transformed into 32 latent variables, concatenated to represent the original distribution for the 64-bar music. 
We attempted this approach, but the results were unsatisfactory.
Since the Multi-view MidiVAE that we used can generate 8-bar segments, our first experiment involved unconditional generation by splitting 16-bar music into two 8-bar segments. 
The results reveal significant content changes between segments, particularly in terms of instrument changes. 
Adding text descriptions as control conditions failed to fully resolve this issue. 
Furthermore, a conditional generation experiment with 32-bar music divided into four 8-bar segments indicated that increasing the number of segments exacerbated the problem of abrupt content changes. 

This problem did not arise in previous work \cite{mittal2021symbolic} because musical information like instrument was not considered. 
Suppose that a musical piece $x$ is segmented into $k$ parts and they are encoded to obtain $k$ distributions $ P(x_1), P(x_2), ..., P(x_k)$.
If these distributions are simply concatenated, the diffusion model would struggle to fit this complex new distribution $P(x_1, x_2, ..., x_k)$.
By introducing textual conditional control $y$, the diffusion model fits a more specific distribution $P(x_1, x_2, ..., x_k | y)$, enhancing its ability to capture instrument characteristics, thus alleviating this issue to some extent.
However, as the number of segments $k$ increases, the complexity of the overall distribution also escalates. 
This leads to an increase of bias in the denoising process of the diffusion model, thereby intensifying this challenge.

To enhance the duration of music generated, we propose a novel autoregressive approach based on music context information. 
Unlike~\cite{mittal2021symbolic}, we generate only one music segment at a time and extend the total length of the music by concatenating these segments.
To ensure the quality of the concatenated music, it is crucial to maintain consistency in music attributes such as instruments, melody, and tempo across all segments.
Therefore, we use the content information from the musical context to guide the generation process of new segments. 
Specifically, for two adjacent music segments, $x^\prime$ and $x$, we use a VAE encoder to extract their latent variables $z^\prime$ and $z$, where $z^\prime$ serves as the contextual content condition $C_{cont}$ for $z$. 
To incorporate $z^\prime$, we design a musical contextual module, which consists of cross-attention layers in each Transformer block.
During training, $z^\prime$ is used as the key and value of the cross-attention layers in the contextual module.
\begin{figure}[ht]
  \centering
\includegraphics[width=1.0\linewidth]{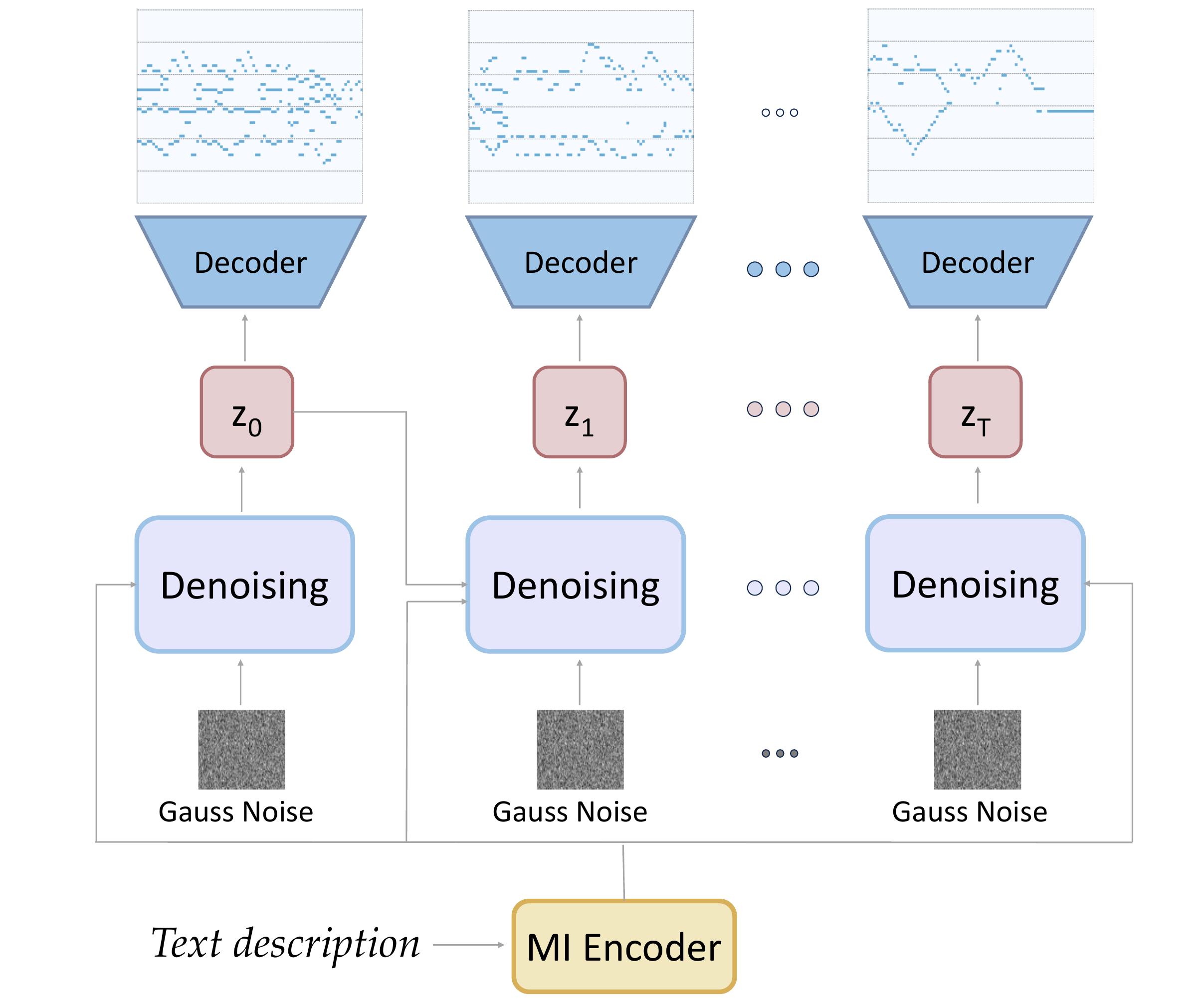}
  \caption{The inference progress for long-duration music generation.}
\label{fig:inference}
\end{figure}
Since the LDM has been trained to fit the distribution $P(z|t, C_{txt})$ and can generate symbolic music matching textual descriptions, we perform full-parameter fine-tuning to enable the model to fit the distribution $P(z|t, C_{txt}, C_{cont})$. 
This allows the model to generate symbolic music that aligns with both the textual descriptions and the musical context.
During fine-tuning, the goal of conditional LDM is changed to mirror score matching by optimizing the
denoising objective.
\begin{equation}
    \mathcal{L}_{\text{cond}} = \mathbb{E}_{z_0, t, \epsilon} \left\| \epsilon - \epsilon_{\theta}(z_t, t, C_{txt}, C_{cont}) \right\|^2
\end{equation}

As shown in Figure \ref{fig:inference}, we first use the LDM without fine-tuning to generate music latent variable$z_0$ that matches the textual descriptions.
The variable $z_0$ is then used as contextual condition for the LDM with fine-tuning, guiding the generation of new music latent variable $z_1$ with the text condition.
Its denoising process is as follows.
\begin{equation}
    p_{\theta}(z_{t-1} \mid z_t) = \mathcal{N} \left(z_{t-1}; \mu_{\theta}(z_t, t, C_{txt}, C_{cont}), \sigma^2_t I \right)
\end{equation}
{\small
\begin{equation}
    \mu_{\theta}(z_t, t, C_{txt}, C_{cont}) = \frac{1}{\sqrt{\alpha_t}} \left( z_t - \frac{1 - \alpha_t}{\sqrt{1 - \overline{\alpha}_t}} \tilde{\epsilon}_{\theta}(z_t, t, C_{txt}, C_{cont}) \right)
\end{equation}

\begin{equation}
    \tilde{\epsilon}_{\theta}(z_t, t, C_{txt},C_{cont}) \leftarrow \omega \epsilon_{\theta}(z_t, t, C_{txt}, C_{cont}) + (1 - \omega) \epsilon_{\theta}(z_t, t)
\end{equation}
}
This generation process can continue until the model has generated the symbolic music of a target length. 
Through this iterative method, we are able to produce multiple music latent variable $\{z_0, z_1, ..., z_T\}$ with consistent musical properties. 
Finally, by decoding these variables and concatenating the segments together in temporal order, we can create high-quality, long-duration music.
\subsection{Music Continuation}
It is worth noting that the proposed autoregressive LDM also enables Diff-Symbo to continue composing music given an existing music piece. 
Specifically, we can use VAE encoder to obtain the latent representation of the original music as the contextual condition. 
Moreover, as discussed in data construction, we can construct a textual description for the desired music by combining musical attributes and textual templates. 
This way, Diff-Symbo can be guided to continue composing the music with both contextual and textual conditions.

\section{Experiments}
\subsection{Experiment Setup}
\textbf{Datasets} 
To train Diff-Symbo, we combined The Lakh MIDI Dataset \cite{raffel2016learning}, EMOPIA \cite{hung2021emopia}, POP909 \cite{wang2020pop909} and Symphony \cite{liu2022symphony} as our MIDI dataset, which contain 224,928 MIDI files.
To train the LDM, we cut each MIDI file into 8-bar music segments and discarded those music segments that do not contain notes within a single bar to ensure the data quality.
This way, we obtained a total of 643,293 8-bar music segments.
For the fine-tuning of the LDM, we cut each MIDI file into 16-bar 
music segments, and then removed those segments where the musical attributes (e.g. instrument) are inconsistent between two consecutive 8-bar music segments. 
This way, we obtained 256,154 16-bar music segments that are musically consistent.  
Then, we extracted music attributes from the music segments and constructed text descriptions by employing the text templates created in Section 3.1.
In real applications, the text input given by the user may not include all the music attributes, especially the less commonly used attributes such as \textit{rhythm} and \textit{danceability}. 
Therefore, during training, we randomly selected 3 to 5 attributes, and for each attribute, we removed it with a probability of 5\% to simulate this situation.
This way, the dataset can better mimic real-world application scenarios and help to train a more robust model.
For both datasets used for LDM training and fine-turning, We divided them into training, validation and test sets in a ratio of 96:2:2.

\textbf{Evaluation Metrics}
To comprehensively evaluate the performance of Diff-Symbo, we conducted a series of objective and subjective experiments with multiple evaluation metrics.
In terms of objective experiments, we followed MuseCoco \cite{lu2023MuseCoco} and Gautam \cite{mittal2021symbolic}, using Average Sample Accuracy (ASA), Fréchet Distance (FD) and Maximum Mean Discrepancy (MMD) as evaluation metrics. 
ASA is used to assess how well generated music matches the corresponding textual descriptions, 
while FD and MMD are used to assess the quality and diversity of the generated samples by quantifying the similarity between the data distributions of model-generated music samples and original music samples in the latent space.  
For subjective evaluation, we invited 20 music enthusiasts to rate the generated samples in three aspects: \textit{melody}, \textit{overall quality} and \textit{controllability}. 
Each metric was rated on a scale of 1-5, corresponding to bad, poor, fair, good and excellent.
Through this method, we obtained the mean opinion score (MOS) results, which served as a measure of subjective perception.
The \textit{controllability}, similar to ASA, is used to measure the match of the generated music to the textual description, while \textit{melody} and \textit{quality} are used to assess the artistry and overall quality of the music.

Additionally, to validate the effectiveness of contextual strategy and the autoregressive approach, we conducted experiments in music continuation and long-duration music generation, respectively. 
For the music continuation experiment, we focused on the consistency of the continuation music with the original music in terms of melody, using \textit{consistency} as an evaluation metric. 
In addition, we utilize \textit{quality} to evaluate the quality of the continuation music.
For the long-duration music generation experiment, we focused on the overall quality of the whole generated music, using \textit{melody} and \textit{quality} as evaluation metrics. 
For both experiments, the music was concatenated by music segments, so we used \textit{coherence} to measure the smoothness at joints. 

\begin{table*}[ht]
    \caption{The MOS results on with 95\% confidence intervals and the result of ASA, FD and MMD of different models.  }
    \vspace{-0.15cm}
    \centering
    \begin{tabular}{lcccccc}
        \toprule
        Model & \textbf{Melody} $\uparrow$ &  \textbf{Controllability} $\uparrow$ & \textbf{Quality} $\uparrow$ & \textbf{ASA (\%)} $\uparrow$& FD $\downarrow$ & MMD $\downarrow$ \\
        \midrule
        GPT-4 & $3.28 \pm 0.09$ & $3.16 \pm 0.10$ & $3.15 \pm 0.08$ & 65.19 & - & - \\
        MuseCoco & $3.32 \pm 0.09$ & $3.22 \pm 0.10$ & $3.30 \pm 0.09$ & 74.89 & 138.93 & 32.83 \\
        Ours & $ \textbf{3.59} \pm \textbf{0.09}$ & $\textbf{3.42} \pm \textbf{0.09}$ & $ \textbf{3.54} \pm \textbf{0.09}$ & \textbf{83.15} & \textbf{93.67} & \textbf{3.05} \\
        \bottomrule
    \end{tabular}
    \label{tab:8-bar}
\end{table*}
\begin{table*}[ht]
    \caption{The ablation experiments measured in terms of the MOS results with 95\% confidence intervals and the result of ASA, FD and MMD.}
    \vspace{-0.2cm}
    \centering
    \scalebox{1.0}{
    \begin{tabular}{lcccccc}
        \toprule
        Model & \textbf{Melody} $\uparrow$ & \textbf{Controllability} $\uparrow$ & \textbf{Quality} $\uparrow$  & \textbf{ASA (\%)} $\uparrow$ & \textbf{FD} $\downarrow$ & \textbf{MMD} $\downarrow$\\
        \midrule
        Ours & $ \textbf{3.59} \pm \textbf{0.09}$ & $3.42 \pm 0.09$ & $ \textbf{3.54} \pm \textbf{0.09}$ &
        83.15 & \textbf{93.67} & \textbf{3.05} \\
        w/ BERT & - & - & -  &
        80.45 & - & - \\
        w/ CFG & $3.45 \pm 0.09$ & $\textbf{3.70} \pm \textbf{0.10}$ & $3.40 \pm 0.09$ & \textbf{86.69} & 115.1 & 5.41 \\
        \bottomrule
    \end{tabular}
}
    \label{tab:ablation}
    \vspace{-0.1cm}
\end{table*}

\textbf{Baselines}
There do not exist many text-to-symbolic music generation works, from which we chose two models that align with the task of this paper as our baselines: GPT-4 \cite{achiam2023gpt} and MuseCoco \cite{lu2023MuseCoco}.
Additionally, to verify the long-duration generation capability of Diff-Symbo, we also conducted a comparison with GPT-4 \cite{achiam2023gpt} and MMT \cite{dong2023multitrack}.
Although MMT cannot generate music based on text, it can produce long-duration music according to instrument categories and also supports music continuation.
To use GPT-4 for symbolic music generation, we followed the strategy of MuseCoco that we instructed GPT-4 to generate symbolic music with 20 texts randomly selected from the test set using the official web page manually. 
To conduct a fair comparison, we trained MuseCoco and MMT using our dataset. Note that we could not conduct music continutation and long-duration music generation experiments with MuseCoco due to its lack of capability in these two aspects.

\textbf{System Configuration} 
The backbone model of our LDM consists of 10 Transformer blocks with 8 attention heads.
The hidden size of attention layers and Feed-Forward Neural Network is 512 and 2048, respectively.
Both training and fine-tuning of the LDM, we set the batch size to 64 and the learning rate to $1 \times 10^{-4}$ with Adam optimizer.
The trainable parameter of the model without fine-tuned is 90M, and we trained it for 48 hours using a single 40G-A100. The trainable parameter of the model with fine-tuned is 108M, and we trained it for 10 hours using a single 40G-A100.

\subsection{Text-to-8-Bar Music Generation}
We first evaluated the model capability of text-to-symbolic music generation and reported the results in Table~\ref{tab:8-bar}. 
The results show that Diff-Symbo has achieved the best performance on all objective and subjective metrics. 
Specifically, the ASA results indicate that our method has superior text controllability than baselines.
Additionally, the FD and MMD results reveal that our model generates music with greater diversity.
In addition, our model also performs better than the baselines in all subjective aspects. 
In subjective evaluations, our model consistently excels over baselines, particularly in terms of \textit{melody} and \textit{quality}, demonstrating improved artistry and musicality. 
Furthermore, our method's superior text controllability is reaffirmed by the results of \textit{controllability} assessments.
These results show that our model can generate symbolic music with better alignment with textual description, higher musicality, more harmonious melody, and overall better quality.

\begin{table}[ht]
    \caption{The MOS results of music continuation with 95\% confidence intervals. GT represents ground truth.}
    \vspace{-0.15cm}
    \centering
    \begin{tabular}{lccc}
        \toprule
         & \textbf{Consistency} $\uparrow$ & \textbf{Coherence} $\uparrow$ & \textbf{Quality} $\uparrow$ \\
        \midrule
        GT & $\textbf{4.01} \pm \textbf{0.11}$ & $\textbf{4.06} \pm \textbf{0.10}$ & $\textbf{3.99} \pm \textbf{0.09}$ \\
        \midrule
        GPT-4 & $3.07 \pm 0.13$ & $3.23 \pm 0.13$ & $3.14 \pm 0.12$ \\
        MMT & $3.14 \pm 0.12$ & $3.31 \pm 0.11$ & $3.28 \pm 0.12$ \\
        Ours & $3.69 \pm 0.09$ & $3.84 \pm 0.09$ & $3.67 \pm 0.10$ \\
        \bottomrule
    \end{tabular}
    \label{tab:music_continuation}
\end{table}

\begin{table}[ht]
    \caption{The MOS results of text-to-32-bar music generation with 95\% confidence intervals.}
    \vspace{-0.15cm}
    \centering
    \begin{tabular}{lccc}
        \toprule
        & \textbf{Melody} $\uparrow$ & \textbf{Coherence} $\uparrow$ & \textbf{Quality} $\uparrow$ \\
        \midrule
        GPT-4 & $2.90 \pm 0.09$ & $3.58 \pm 0.08$ & $2.94 \pm 0.08$ \\
        MMT & $3.12 \pm 0.11$ & $3.32 \pm 0.08$ & $3.15 \pm 0.09$ \\
        Ours & $\textbf{3.74} \pm \textbf{0.08}$ & $\textbf{3.86} \pm \textbf{0.07}$ & $\textbf{3.66} \pm \textbf{0.09}$ \\
        \bottomrule
    \end{tabular}
    \label{tab:32-bars}
    \vspace{-0.3cm}
\end{table}

\subsection{Music Continuation}
For the music continuation experiment, we provided an 8-bar segment of original music and required each model to continue with an additional 8-bar music.
We implemented the contextual learning strategy to accomplish the experiment, and the results in Table~\ref{tab:music_continuation} validate its effectiveness. 
Specifically, the \textit{consistency} score significantly surpasses that of GPT-4 and MMT, closely following the ground truth music pieces.
This demonstrates that our contextual learning strategy effectively maintains consistency in rhythm and melody between the extended and original music
The \textit{coherence} score confirms that our model can achieve smooth transitions when extending music, ensuring the effectiveness of our concatenation method. 
Additionally, the \textit{quality }score indicates that the music extended by our model has better artistic and musical qualities.
These results show the effectiveness of our contextual learning strategy as well as the model's music continuation ability.


\subsection{Text-to-32-Bar Music Generation}
To validate the capability of our model in generating long-duration music, we conducted a text-to-32-bar music generation experiment, typically lasting over one minute.
The \textit{quality} scores in Table~\ref{tab:32-bars} show that the 32-bar music generated by our model significantly surpasses that of GPT-4 and MMT. 
Additionally, the scores of \textit{melody} and \textit{coherence} show that despite being concatenated by multiple segments, our long-duration music maintain consistency in melody and rhythm with smooth transitions between segments due to the effective contextual learning strategy.
These results demonstrate the effectiveness of our contextual learning strategy and our model's ability to generate high-quality, long-duration music.
This is crucial for creating song-level music.
\vspace{-0.2cm}
\section{Ablation Study}
\textbf{MI Encoder}
In order to check that the features extracted from music information encoder are better than the pre-trained BERT in focusing on musical attributes in textual descriptions, we conducted an experiment using the ASA metric.
As shown in Table \ref{tab:ablation}, there is an improvement in the ASA results for the model with fine-tuning of BERT, which demonstrates that fine-tuning BERT is able to better capture the features related to music attributes within the text description, resulting in a higher degree of consistency between the generated music and the text description.

\textbf{CFG}
We studied the impact of CFG with scale $\omega = 7.5$. As shown in Table \ref{tab:ablation}, the use of CFG during inference can improve the textual controllability of the music generation but slightly degrade the music quality. 
The usage of CFG enhances the model's responsiveness to textual descriptions, improving the textual control over the generated music. 
However, this might concurrently limit the model's exploratory space during the generation process, potentially compromising the overall quality of the music \cite{ho2022classifier}. 

\vspace{-0.2cm}
\section{Conclusion}
In this paper, we presented Diff-Symbo, a model for generating high-quality, long-duration symbolic
music that matches text description. To the best of our knowledge, this is the first work to apply LDM
and contextual learning to multi-track text-to-symbolic music generation. 
In addition, to address the problem of insufficient datasets, we constructed a larger and more comprehensive dataset of text templates. By employing an contextual learning strategy, we enabled the model to generate long-duration music with higher coherence and better melody. 
Experiment results demonstrated that our model has a significant advantage over other baselines in generating high-quality, long-duration music.

\appendix
\bibliography{aaai25}

@article{lu2023musecoco,
  title={Musecoco: Generating symbolic music from text},
  author={Lu, Peiling and Xu, Xin and Kang, Chenfei and Yu, Botao and Xing, Chengyi and Tan, Xu and Bian, Jiang},
  journal={arXiv preprint arXiv:2306.00110},
  year={2023}
}

@article{achiam2023gpt,
  title={Gpt-4 technical report},
  author={Achiam, Josh and Adler, Steven and Agarwal, Sandhini and Ahmad, Lama and Akkaya, Ilge and Aleman, Florencia Leoni and Almeida, Diogo and Altenschmidt, Janko and Altman, Sam and Anadkat, Shyamal and others},
  journal={arXiv preprint arXiv:2303.08774},
  year={2023}
}

@article{schneider2023mo,
  title={Mo$\backslash$\^{} usai: Text-to-music generation with long-context latent diffusion},
  author={Schneider, Flavio and Kamal, Ojasv and Jin, Zhijing and Sch{\"o}lkopf, Bernhard},
  journal={arXiv preprint arXiv:2301.11757},
  year={2023}
}

@article{huang2022mulan,
  title={Mulan: A joint embedding of music audio and natural language},
  author={Huang, Qingqing and Jansen, Aren and Lee, Joonseok and Ganti, Ravi and Li, Judith Yue and Ellis, Daniel PW},
  journal={arXiv preprint arXiv:2208.12415},
  year={2022}
}

@inproceedings{zhang2020butter,
  title={BUTTER: A representation learning framework for bi-directional music-sentence retrieval and generation},
  author={Zhang, Yixiao and Wang, Ziyu and Wang, Dingsu and Xia, Gus},
  booktitle={Proceedings of the 1st workshop on nlp for music and audio (nlp4musa)},
  pages={54--58},
  year={2020}
}

@inproceedings{rombach2022high,
  title={High-resolution image synthesis with latent diffusion models},
  author={Rombach, Robin and Blattmann, Andreas and Lorenz, Dominik and Esser, Patrick and Ommer, Bj{\"o}rn},
  booktitle={Proceedings of the IEEE/CVF conference on computer vision and pattern recognition},
  pages={10684--10695},
  year={2022}
}

@inproceedings{ni2023conditional,
  title={Conditional image-to-video generation with latent flow diffusion models},
  author={Ni, Haomiao and Shi, Changhao and Li, Kai and Huang, Sharon X and Min, Martin Renqiang},
  booktitle={Proceedings of the IEEE/CVF Conference on Computer Vision and Pattern Recognition},
  pages={18444--18455},
  year={2023}
}

@article{blattmann2023stable,
  title={Stable video diffusion: Scaling latent video diffusion models to large datasets},
  author={Blattmann, Andreas and Dockhorn, Tim and Kulal, Sumith and Mendelevitch, Daniel and Kilian, Maciej and Lorenz, Dominik and Levi, Yam and English, Zion and Voleti, Vikram and Letts, Adam and others},
  journal={arXiv preprint arXiv:2311.15127},
  year={2023}
}

@article{bar2024lumiere,
  title={Lumiere: A space-time diffusion model for video generation},
  author={Bar-Tal, Omer and Chefer, Hila and Tov, Omer and Herrmann, Charles and Paiss, Roni and Zada, Shiran and Ephrat, Ariel and Hur, Junhwa and Li, Yuanzhen and Michaeli, Tomer and others},
  journal={arXiv preprint arXiv:2401.12945},
  year={2024}
}

@article{podell2023sdxl,
  title={Sdxl: Improving latent diffusion models for high-resolution image synthesis},
  author={Podell, Dustin and English, Zion and Lacey, Kyle and Blattmann, Andreas and Dockhorn, Tim and M{\"u}ller, Jonas and Penna, Joe and Rombach, Robin},
  journal={arXiv preprint arXiv:2307.01952},
  year={2023}
}

@article{muller2023multimodal,
  title={A multimodal comparison of latent denoising diffusion probabilistic models and generative adversarial networks for medical image synthesis},
  author={M{\"u}ller-Franzes, Gustav and Niehues, Jan Moritz and Khader, Firas and Arasteh, Soroosh Tayebi and Haarburger, Christoph and Kuhl, Christiane and Wang, Tianci and Han, Tianyu and Nolte, Teresa and Nebelung, Sven and others},
  journal={Scientific Reports},
  volume={13},
  number={1},
  pages={12098},
  year={2023},
  publisher={Nature Publishing Group UK London}
}

@article{liu2023audioldm,
  title={Audioldm: Text-to-audio generation with latent diffusion models},
  author={Liu, Haohe and Chen, Zehua and Yuan, Yi and Mei, Xinhao and Liu, Xubo and Mandic, Danilo and Wang, Wenwu and Plumbley, Mark D},
  journal={arXiv preprint arXiv:2301.12503},
  year={2023}
}

@article{ghosal2023text,
  title={Text-to-audio generation using instruction-tuned llm and latent diffusion model},
  author={Ghosal, Deepanway and Majumder, Navonil and Mehrish, Ambuj and Poria, Soujanya},
  journal={arXiv preprint arXiv:2304.13731},
  year={2023}
}

@inproceedings{chen2024musicldm,
  title={MusicLDM: Enhancing novelty in text-to-music generation using beat-synchronous mixup strategies},
  author={Chen, Ke and Wu, Yusong and Liu, Haohe and Nezhurina, Marianna and Berg-Kirkpatrick, Taylor and Dubnov, Shlomo},
  booktitle={ICASSP 2024-2024 IEEE International Conference on Acoustics, Speech and Signal Processing (ICASSP)},
  pages={1206--1210},
  year={2024},
  organization={IEEE}
}

@article{mittal2021symbolic,
  title={Symbolic music generation with diffusion models},
  author={Mittal, Gautam and Engel, Jesse and Hawthorne, Curtis and Simon, Ian},
  journal={arXiv preprint arXiv:2103.16091},
  year={2021}
}

@article{liu2023audioldm2,
  title={AudioLDM 2: Learning holistic audio generation with self-supervised pretraining},
  author={Liu, Haohe and Tian, Qiao and Yuan, Yi and Liu, Xubo and Mei, Xinhao and Kong, Qiuqiang and Wang, Yuping and Wang, Wenwu and Wang, Yuxuan and Plumbley, Mark D},
  journal={arXiv preprint arXiv:2308.05734},
  year={2023}
}

@inproceedings{roberts2018hierarchical,
  title={A hierarchical latent vector model for learning long-term structure in music},
  author={Roberts, Adam and Engel, Jesse and Raffel, Colin and Hawthorne, Curtis and Eck, Douglas},
  booktitle={International conference on machine learning},
  pages={4364--4373},
  year={2018},
  organization={PMLR}
}

@article{melechovsky2023mustango,
  title={Mustango: Toward controllable text-to-music generation},
  author={Melechovsky, Jan and Guo, Zixun and Ghosal, Deepanway and Majumder, Navonil and Herremans, Dorien and Poria, Soujanya},
  journal={arXiv preprint arXiv:2311.08355},
  year={2023}
}

@article{lin2024multi,
  title={Multi-view MidiVAE: Fusing Track-and Bar-view Representations for Long Multi-track Symbolic Music Generation},
  author={Lin, Zhiwei and Chen, Jun and Tang, Boshi and Sha, Binzhu and Yang, Jing and Ju, Yaolong and Fan, Fan and Kang, Shiyin and Wu, Zhiyong and Meng, Helen},
  journal={arXiv preprint arXiv:2401.07532},
  year={2024}
}

@article{devlin2018bert,
  title={Bert: Pre-training of deep bidirectional transformers for language understanding},
  author={Devlin, Jacob and Chang, Ming-Wei and Lee, Kenton and Toutanova, Kristina},
  journal={arXiv preprint arXiv:1810.04805},
  year={2018}
}

@article{copet2024simple,
  title={Simple and controllable music generation},
  author={Copet, Jade and Kreuk, Felix and Gat, Itai and Remez, Tal and Kant, David and Synnaeve, Gabriel and Adi, Yossi and D{\'e}fossez, Alexandre},
  journal={Advances in Neural Information Processing Systems},
  volume={36},
  year={2024}
}

@article{kingma2013auto,
  title={Auto-encoding variational bayes},
  author={Kingma, Diederik P and Welling, Max},
  journal={arXiv preprint arXiv:1312.6114},
  year={2013}
}

@article{ho2020denoising,
  title={Denoising diffusion probabilistic models},
  author={Ho, Jonathan and Jain, Ajay and Abbeel, Pieter},
  journal={Advances in neural information processing systems},
  volume={33},
  pages={6840--6851},
  year={2020}
}

@article{ho2022classifier,
  title={Classifier-free diffusion guidance},
  author={Ho, Jonathan and Salimans, Tim},
  journal={arXiv preprint arXiv:2207.12598},
  year={2022}
}

@article{raffel2020exploring,
  title={Exploring the limits of transfer learning with a unified text-to-text transformer},
  author={Raffel, Colin and Shazeer, Noam and Roberts, Adam and Lee, Katherine and Narang, Sharan and Matena, Michael and Zhou, Yanqi and Li, Wei and Liu, Peter J},
  journal={Journal of machine learning research},
  volume={21},
  number={140},
  pages={1--67},
  year={2020}
}

@inproceedings{elizalde2023clap,
  title={Clap learning audio concepts from natural language supervision},
  author={Elizalde, Benjamin and Deshmukh, Soham and Al Ismail, Mahmoud and Wang, Huaming},
  booktitle={ICASSP 2023-2023 IEEE International Conference on Acoustics, Speech and Signal Processing (ICASSP)},
  pages={1--5},
  year={2023},
  organization={IEEE}
}

@article{van2017neural,
  title={Neural discrete representation learning},
  author={Van Den Oord, Aaron and Vinyals, Oriol and others},
  journal={Advances in neural information processing systems},
  volume={30},
  year={2017}
}

@article{evans2024fast,
  title={Fast Timing-Conditioned Latent Audio Diffusion},
  author={Evans, Zach and Carr, CJ and Taylor, Josiah and Hawley, Scott H and Pons, Jordi},
  journal={arXiv preprint arXiv:2402.04825},
  year={2024}
}

@article{evans2024long,
  title={Long-form music generation with latent diffusion},
  author={Evans, Zach and Parker, Julian D and Carr, CJ and Zukowski, Zack and Taylor, Josiah and Pons, Jordi},
  journal={arXiv preprint arXiv:2404.10301},
  year={2024}
}

@article{zeghidour2021soundstream,
  title={Soundstream: An end-to-end neural audio codec},
  author={Zeghidour, Neil and Luebs, Alejandro and Omran, Ahmed and Skoglund, Jan and Tagliasacchi, Marco},
  journal={IEEE/ACM Transactions on Audio, Speech, and Language Processing},
  volume={30},
  pages={495--507},
  year={2021},
  publisher={IEEE}
}

@article{defossez2022high,
  title={High fidelity neural audio compression},
  author={D{\'e}fossez, Alexandre and Copet, Jade and Synnaeve, Gabriel and Adi, Yossi},
  journal={arXiv preprint arXiv:2210.13438},
  year={2022}
}

@article{ziv2024masked,
  title={Masked Audio Generation using a Single Non-Autoregressive Transformer},
  author={Ziv, Alon and Gat, Itai and Lan, Gael Le and Remez, Tal and Kreuk, Felix and D{\'e}fossez, Alexandre and Copet, Jade and Synnaeve, Gabriel and Adi, Yossi},
  journal={arXiv preprint arXiv:2401.04577},
  year={2024}
}

@article{agostinelli2023musiclm,
  title={Musiclm: Generating music from text},
  author={Agostinelli, Andrea and Denk, Timo I and Borsos, Zal{\'a}n and Engel, Jesse and Verzetti, Mauro and Caillon, Antoine and Huang, Qingqing and Jansen, Aren and Roberts, Adam and Tagliasacchi, Marco and others},
  journal={arXiv preprint arXiv:2301.11325},
  year={2023}
}

@article{hung2021emopia,
  title={EMOPIA: A multi-modal pop piano dataset for emotion recognition and emotion-based music generation},
  author={Hung, Hsiao-Tzu and Ching, Joann and Doh, Seungheon and Kim, Nabin and Nam, Juhan and Yang, Yi-Hsuan},
  journal={arXiv preprint arXiv:2108.01374},
  year={2021}
}

@article{wang2020pop909,
  title={Pop909: A pop-song dataset for music arrangement generation},
  author={Wang, Ziyu and Chen, Ke and Jiang, Junyan and Zhang, Yiyi and Xu, Maoran and Dai, Shuqi and Gu, Xianbin and Xia, Gus},
  journal={arXiv preprint arXiv:2008.07142},
  year={2020}
}

@book{raffel2016learning,
  title={Learning-based methods for comparing sequences, with applications to audio-to-midi alignment and matching},
  author={Raffel, Colin},
  year={2016},
  publisher={Columbia University}
}

@article{liu2022symphony,
  title={Symphony generation with permutation invariant language model},
  author={Liu, Jiafeng and Dong, Yuanliang and Cheng, Zehua and Zhang, Xinran and Li, Xiaobing and Yu, Feng and Sun, Maosong},
  journal={arXiv preprint arXiv:2205.05448},
  year={2022}
}

@article{wu2022exploring,
  title={Exploring the efficacy of pre-trained checkpoints in text-to-music generation task},
  author={Wu, Shangda and Sun, Maosong},
  journal={arXiv preprint arXiv:2211.11216},
  year={2022}
}

@inproceedings{dong2023multitrack,
  title={Multitrack music transformer},
  author={Dong, Hao-Wen and Chen, Ke and Dubnov, Shlomo and McAuley, Julian and Berg-Kirkpatrick, Taylor},
  booktitle={ICASSP 2023-2023 IEEE International Conference on Acoustics, Speech and Signal Processing (ICASSP)},
  pages={1--5},
  year={2023},
  organization={IEEE}
}

@article{min2023polyffusion,
  title={Polyffusion: A diffusion model for polyphonic score generation with internal and external controls},
  author={Min, Lejun and Jiang, Junyan and Xia, Gus and Zhao, Jingwei},
  journal={arXiv preprint arXiv:2307.10304},
  year={2023}
}

@article{wang2024whole,
  title={Whole-song hierarchical generation of symbolic music using cascaded diffusion models},
  author={Wang, Ziyu and Min, Lejun and Xia, Gus},
  journal={arXiv preprint arXiv:2405.09901},
  year={2024}
}

\end{document}